\documentclass[11pt,twoside]{article}

\usepackage{asp2006}
\usepackage{epsf}
\usepackage{lscape}
\usepackage{graphicx}

\begin{document}

\title{FUV Spectroscopy of the sdOB Primary of the Eclipsing Binary System AA Dor}   

\author{Johannes Fleig$^1$, Thomas Rauch$^1$, Klaus Werner$^1$, Jeffrey W\@. Kruk$^2$}   

\affil{
$^1$Institute for Astronomy and Astrophysic,
Kepler Center for Astro and Particle Physics,
Eberhard Karls University, T\"ubingen, Germany\\ 
$^2$Department of Physics and Astronomy, Johns Hopkins University, Baltimore, MD 21218, USA}   

\begin{abstract}
AA~Dor is an eclipsing, close, post common-envelope binary (PCEB).
We present a detailed spectral analysis of its sdOB primary star
based on observations obtained with the Far Ultraviolet
Spectroscopic Explorer (FUSE). Due to a strong contamination by
interstellar absorption, we had to model both, the stellar spectrum
as well as the interstellar line absorption in order to reproduce the
FUV observation well and to determine the photospheric parameters
precisely.

 \end{abstract}

\section{Introduction}   
\begin{figure}[ht!]
 \begin{center}
 \includegraphics[width=0.7\textwidth]{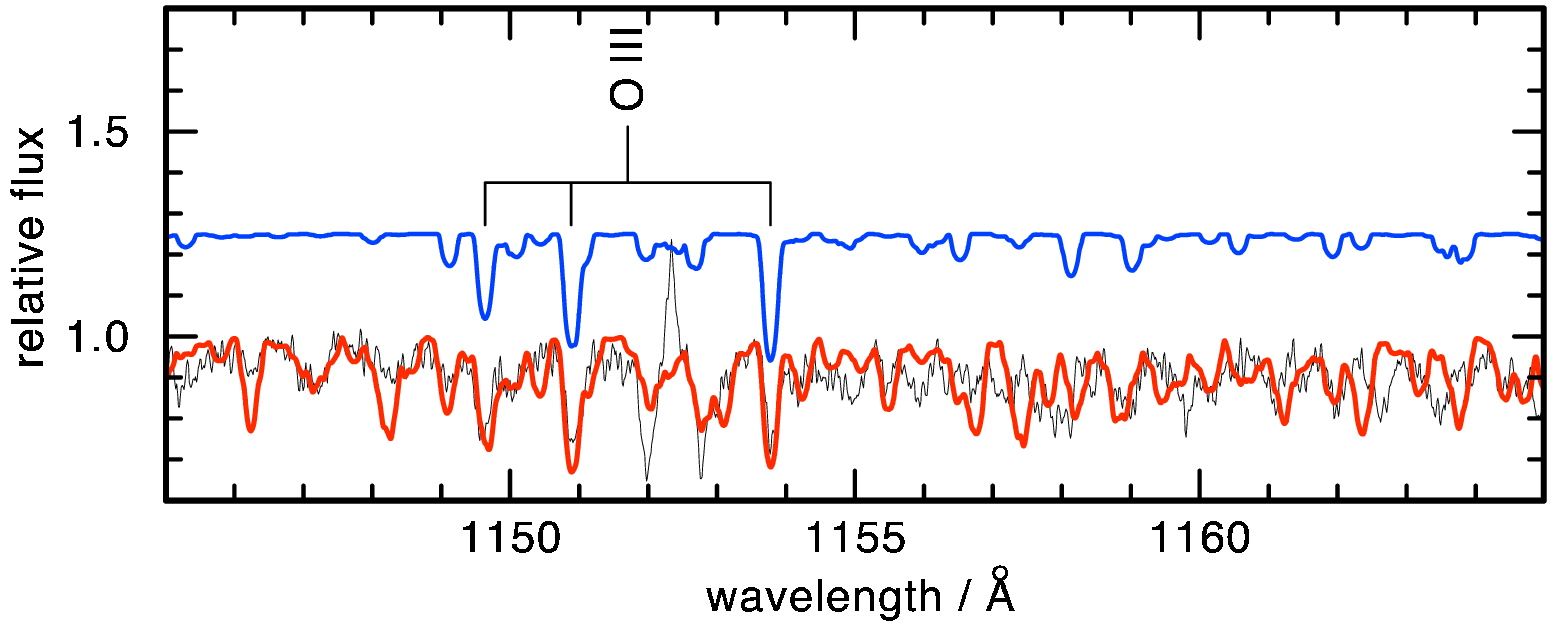}\vspace{-2mm}
 \caption[]{
Identification of iron-group lines. \citet{Kurucz91} gives only a few lines with reliably measured wavelengths (POS lines,
blue, shifted to the top) which makes the identification of isolated lines difficult. The addition of iron-group lines
with theoretical positions (Kurucz's LIN lines, red) improves the overall-fit to the observation.
The synthetic spectra are convolved with a rotational profile of $v_{\rm{rot}}=35\,\rm{km\,s^{-1}}$.}\vspace{-5mm}
  \label{fig:POSLIN}
\end{center}  
\end{figure}

The eclipsing binary system AA~Dor consists of a sdOB with \mbox{$T_{\rm{eff}}=42\,\rm{kK}$} and an unseen low-mass
companion. 
Recent studies of its primary star have shown discrepancies in the determination of the surface gravity $g$. 
\citet{Rauch00} obtained \mbox{$\log g\hspace{-0.5mm} =\hspace{-0.5mm}  5.21$} by a NLTE spectral analysis of 
optical and ultraviolet data. The results of radial-velocity and lightcurve analyses by \citet{Hilditch96,Hilditch03} 
indicate a higher value of \mbox{$\log g\hspace{-0.5mm} =\hspace{-0.5mm}  5.45\,-\,5.51$}.
Twelve high S/N-ratio FUSE observations were performed in August 2003 and June 2004. 
With about 200 secs each, the exposure times were chosen short in order to reduce the smearing by orbital motion. 
The FUV range includes the hydrogen Lyman series which in general is sensitive to changes in $\log g$. 
However, a precise determination of the surface gravity and further stellar parameters is hampered
by strong interstellar absorption lines.
We used our NLTE code TMAP 
\citep[T\"ubingen Model Atmosphere Package, ][]{Werner03} 
to calculate plane-parallel, line-blanketed model atmospheres in radiative, hydrostatic and statistical 
equilibrium.  
We included calcium and all elements of the iron group (Sc, Ti, V, Cr, Mn, Fe, Co, Ni) 
into our calculations. Due to the large amount of transitions, those elements were treated in a statistical 
approach \citep{rd2003}. 
Atomic data were taken from NIST (National Institute of Standards and Technology, 
\texttt{http://physics.nist.gov}), 
the Opacity Project \citep{Seaton94}, and the Iron Project \citep{Hummer93}. 
In case of the iron-group elements, energy levels and oscillator strengths stem from Kurucz's 
line lists \citep{Kurucz91}. Only weak lines of those elements are detected in the FUSE spectra, 
which form absorption troughs due to rotational broadening (Fig. \ref{fig:POSLIN}). 

\section{Conclusions}

\begin{figure}[ht!]
\begin{center}
 \includegraphics[width=0.7\textwidth]{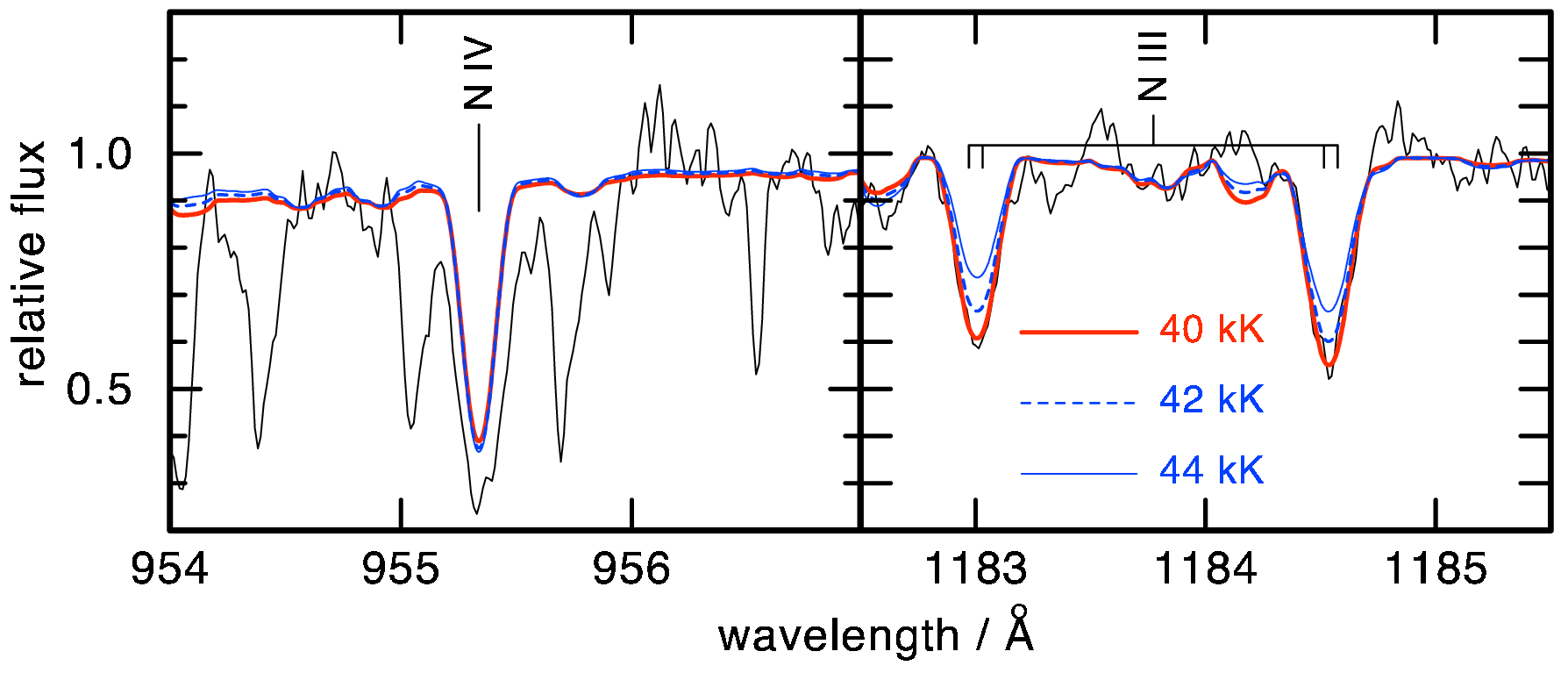}\vspace{-2mm}
 \caption[]{Determination of $T_{\rm{eff}}$ from the 
\mbox{N\,{\scriptsize{III}} / N\,\scriptsize{IV}} ionization equilibrium. While the \mbox{N\,\scriptsize{IV}} 
line (left panel) appears saturated within the relevant $T_{\rm{eff}}$, the 
\mbox{N\,\scriptsize{III}} multiplet (right) matches the observation best at $T_{\rm{eff}} = 40\,\mathrm{kK}$. }
  \label{fig:Teff}
  \end{center}
\end{figure}

\begin{figure}[ht!]
\begin{center}
 \includegraphics[width=0.7\textwidth]{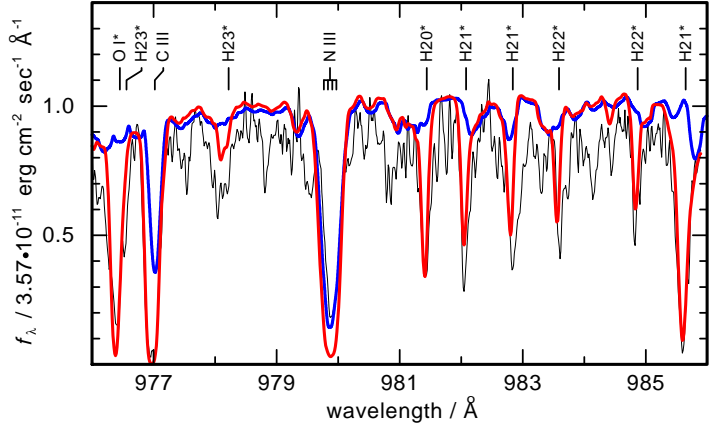}\vspace{-2mm}
  \caption[]{Comparison of a pure photospheric model spectrum (blue) and the combined photospheric
  + ISM model (red) within a section of the FUSE observation. Interstellar lines are labeled with an asterisk.}
  \label{fig:ISM_MOD}
\end{center}
\end{figure}

\begin{figure}[ht!]
\begin{center}
 \includegraphics[width=0.9\textwidth]{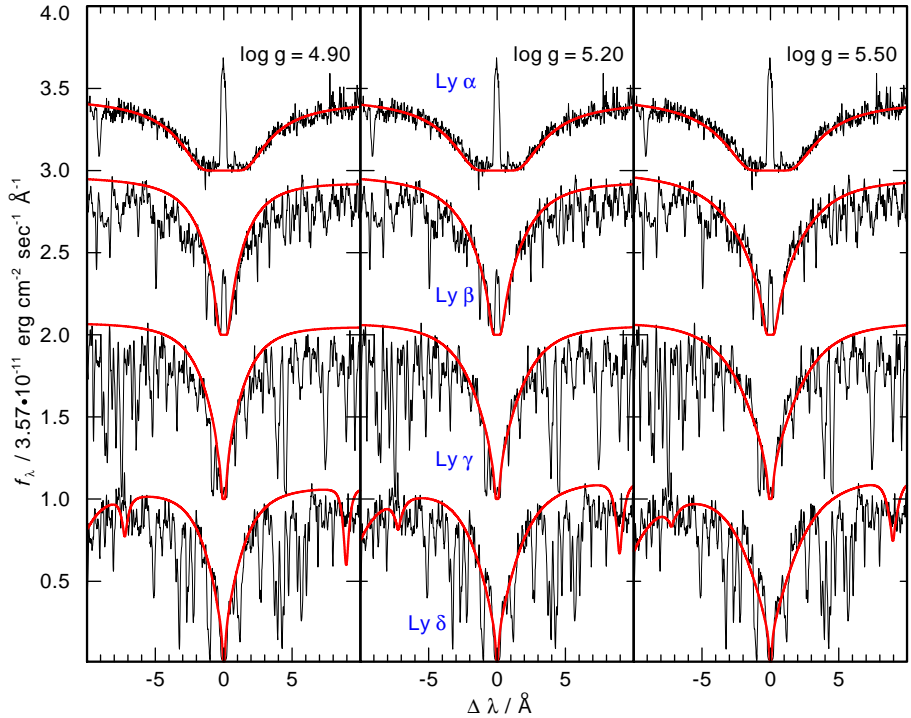}\vspace{-2mm}
 \caption[]{Log $g$ - dependence of the first four lines of the Lyman series. 
            Lyman $\alpha$ is compared to IUE observations, Lyman $\beta - \delta$ to FUSE observations. 
            An interstellar H\,{\scriptsize I} column density of $\log N_\mathrm{H\,{\footnotesize I}} = 20.3$ and an 
            extinction of $E_\mathrm{B-V}=0.04$ are considered.
            Interstellar absorption prevents 
            a precise determination of the continuum and the course of the line wings. 
            All continua are normalized to that of the synthetic spectra with $\log g=5.20$.}
  \label{fig:logg}
  \end{center}
\end{figure}

\begin{figure}[ht!]
\begin{center}
 \includegraphics[width=0.7\textwidth]{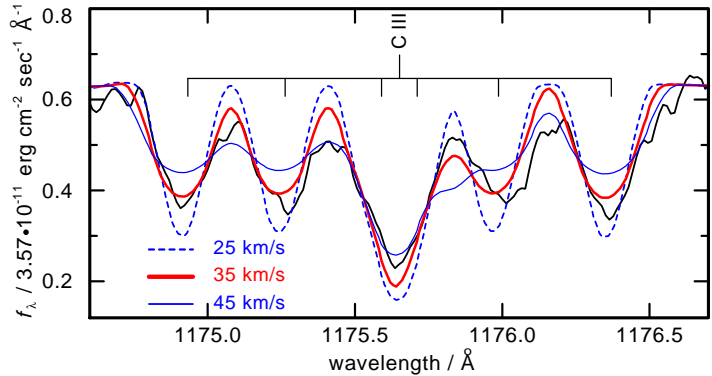}\vspace{-2mm}
  \caption[]{
    Comparison of the \mbox{C\,{\scriptsize III} $\lambda\lambda 1175\,\mathrm{\AA}$} multiplet with
    the FUSE observation at three different rotational velocities $v_{\rm{rot}}$.
    The best fit is achieved with $v_{\rm{rot}}=35\,\rm{km\,s}^{-1}$.}
  \label{fig:vrot}
  \end{center}
\end{figure}

The effective temperature was derived using the ionization equilibrium of N\,{\scriptsize III} and N\,{\scriptsize IV}. 
The best fit is achieved for $T_{\rm{eff}}\hspace{-0.5mm}=\hspace{-0.5mm}40 \pm 3\,\mathrm{kK}$ (Fig\@. \ref{fig:Teff}). 
Due to the interstellar contamination of the spectra (Fig\@. \ref{fig:ISM_MOD}), 
a precise determination of the surface gravity was 
not possible (Fig\@. \ref{fig:logg}). 
However, a hint towards a higher value of $\log g$ is given by the rotational velocity of the primary 
(Fig\@. \ref{fig:vrot}). With $v_{\rm{rot}}=35 \pm 5\,\rm{km\,s}^{-1}$, the primary rotates slower than bound 
($v_{\rm{rot}}=45.7\,\rm{km\,s}^{-1}$). \citet{Rauch00} assumed bound rotation for the calculation of the primary's 
spectrum and thus, one can expect a slightly higher value of $\log g$. Calculating the surface gravity directly via 
$\log\,g=\log\, [(GM_1)/r_1^2]$, $r_1\propto v_{\rm{rot}}\cdot P$, returns a surface gravity of 5.44, which 
is in agreement with the value found by \citet{Hilditch03}. 

\begin{figure}[ht!]
\begin{center}
 \includegraphics[width=0.9\textwidth]{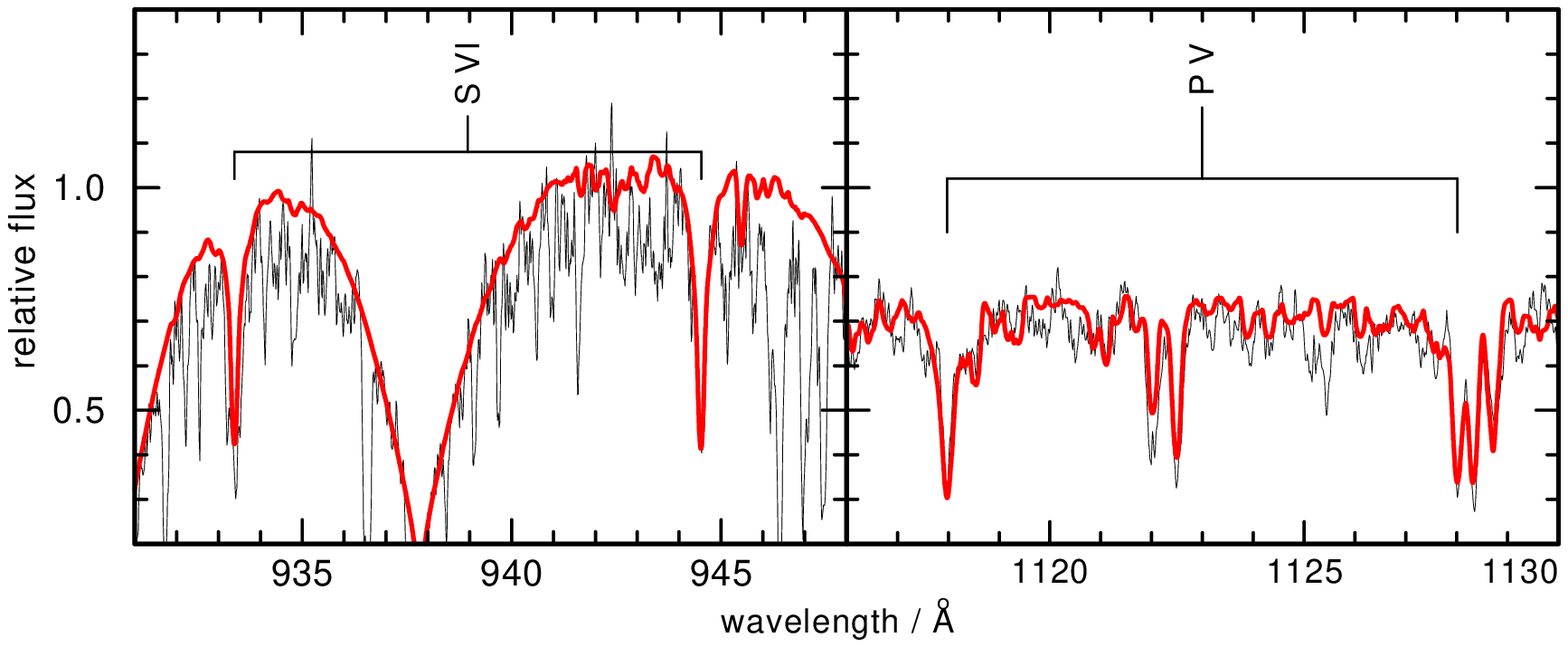}\vspace{-2mm}
 \caption[]{Detection of phosphorus and sulfur. The resonance doublets of 
\mbox{S\,{\scriptsize VI} $\lambda\lambda 933.38, 944.52\,\mathrm{\AA}$} and 
\mbox{P\,{\scriptsize V} $\lambda\lambda 1117.98, 1128.01\,\mathrm{\AA}$} are well reproduced at solar abundances. 
Since interstellar absorption contributions cannot be excluded, these abundances are upper limits.}
 \label{fig:doublets}
  \end{center}
 \end{figure}

\acknowledgements
T.R\@. is supported by the \emph{German Astrophysical Virtual Observatory} project
of the German Federal Ministry of Education and Research (BMBF) under grant 05\,AC6VTB.
This work has been done using the profile fitting procedure {\sc Owens} developed by M\@.
Lemoine and the \mbox{FUSE French Team.}

\end{document}